\documentclass{ws-mpla}
\newcommand{\beq}{\begin{equation}}
\newcommand{\eeq}{\end{equation}}  
\newcommand{\beqn}{\begin{eqnarray}}
\newcommand{\eeqn}{\end{eqnarray}} 
\newcommand{\gppr}{\stackrel{>}{\scriptstyle \sim}}
\newcommand{\lppr}{\stackrel{<}{\scriptstyle \sim}}

\begin{document}

\markboth{F.M. Rieger \& F. Aharonian}{Probing the central black hole in M87}

%%%%%%%%%%%%%%%%%%%%% Publisher's Area please ignore %%%%%%%%%%%%%%
\catchline{}{}{}{}{}
%%%%%%%%%%%%%%%%%%%%%%%%%%%%%%%%%%%%%%%%%%%%%%%%%%%%%%%%%%%%%%%%%%%

\title{Probing the central black hole in M87 with gamma-rays} 

\author{Frank M. Rieger}
\address{Max-Planck-Institut f{\"u}r Kernphysik,\\ 
P.O. Box 103980, 69029 Heidelberg, Germany\\
frank.rieger@mpi-hd.mpg.de}

\author{Felix Aharonian}
\address{Dublin Institute for Advanced Studies,\\ 
31 Fitzwilliam Place, Dublin 2, Ireland\\
felix.aharonian@mpi-hd.mpg.de}

\maketitle

\pub{Received (Day Month Year)}{Revised (Day Month Year)}

\begin{abstract}
Recent high-sensitivity observation of the nearby radio galaxy M87 have provided important insights 
into the central engine that drives the large-scale outflows seen in radio, optical and X-rays. This review 
summarizes the observational status achieved in the high energy (HE;$<$ 100 GeV) and very high energy 
(VHE; $>$100 GeV) gamma-ray domains, and discusses the theoretical progress in understanding the 
physical origin of this emission and its relation to the activity of the central black hole. 
\end{abstract}

\ccode{PACS Nos.: 95.30.Sf, 95.85.Pw, 98.54.Gr, 98.62.Js}

\section{Introduction}	
The advanced capabilities of current gamma-ray instruments offer a unique tool to elucidate the nature of 
the central engine in extragalactic objects. At the time of writing, about 1000 active galaxies have been 
detected above 100 MeV by Fermi-LAT.\cite{ackermann11} Ground-based gamma-ray instruments on 
the other hand have established TeV emission and spectra for about 50 active galaxies.\footnote{For an 
up-to-date list see the online TeV Source Catalog (TeVCat) at http://tevcat.uchicago.edu/} The majority
of these are blazars, only a minor fraction are radio galaxies. The giant elliptical Virgo-cluster galaxy 
M87 (NGC 4486, 3C274) is particularly unique among those sources, being the nearest AGN in the 
northern sky, one of the first extragalactic objects discovered (in 1918) to have an optical jet\cite{curtis1918}, 
the first extragalactic X-ray source to be identified\cite{bradt67}, and the first non-blazar active galaxy 
detected at VHE energies.\cite{aharonian03} Because of its proximity, many astrophysical phenomena 
can be studied in exceptional detail, which has led M87 to become a prominent benchmark for theoretical 
research. In particular, since the jet in M87 is substantially inclined, it is fairly easy to study its core - the 
vicinity of the black hole.

\section{General properties of M87}
(1) Located at a distance of $d \simeq 16.7$ Mpc (redshift of $z=0.004$; 1 mas=0.081 pc),\cite{mei07,bird10} 
M87 is commonly counted among the FR~I class\cite{fanaroff74} of (weak power) radio galaxies. It is 
considered to host one of the most massive black holes (BH) in the Universe with mass estimates based on 
gas or stellar dynamics in the range of $M_{\rm BH} \simeq (2-6) \times 10^9\,M_{\odot}$,\cite{marconi97,tremaine02,gebhardt09}
(see also Ref.~\refcite{hlavacek12} for a fundamental plane inference).\cite{peterson10} 
There is evidence in HST data for a possible offset between the galaxy (bulge) photo-center and the AGN 
that might be related to a gravitational wave recoil resulting from the coalescence of a supermassive BH 
binary.\cite{batcheldor10}\\
(2) M87 is well-known for a bright one-sided, quasi-straight jet (of projected length $\sim2$ kpc) that extends 
from the center to the outer parts of the source and can be clearly seen in radio, optical and X-rays, cf.
Fig.~1.\cite{biretta95,marshall02} On very large scales, there is evidence (VLA, 90 cm) for two directed, 
coherent outflows powering the giant radio halo extending about 40 kpc from the nucleus.\cite{owen00}\\
(3) High-resolution VLBA images suggest that the innermost part of the jet seen at radio frequencies is formed 
very close to the black hole, with an apparent (initial) 'full opening angle' of approximately $\theta_{j,a}\simeq
60^{\circ}$. Significant collimation of the jet already occurs within a few tens of Schwarzschild radii ($r_s$) 
from the black hole, see e.g. Fig.~2, but seems not complete until about $1000 r_s$.\cite{junor99,ly04,walker08,hada11} 
An apparent full opening angle of $60^{\circ}$ (i.e., intrinsic full opening angle $\theta_j \simeq \theta_{j,a}
\sin \theta \simeq 30^{\circ}$ for a jet inclination $\theta=20^{\circ}$, cf. Ref.~\refcite{acciari09Sci}) up to 
projected distances of about $100 r_s$,\cite{biretta02} suggests that the (radio) jet already has a comparable 
width on those scales.\\ 
(4) Despite its huge black hole mass, M87 is not a powerful source of radiation. Common estimates for its 
total nuclear (disk and jet) bolometric luminosity do not exceed $L_{\rm bol} \simeq 10^{42}$ erg/s by 
much.\cite{owen00,biretta91,whysong04} When this is compared with the canonical Eddington luminosity 
value $L_{\rm Edd}$, M87 becomes highly under-luminous by a factor $l_e = L_{\rm bol}/L_{\rm Edd} \leq 
3 \times 10^{-6}$. This has led to the proposal that M87 is a prototype galaxy, where accretion occurs in 
a non-standard, advective-dominated (ADAF) mode characterized by an intrinsically low radiative 
efficiency.\cite{reynolds96,camenzind99,dimatteo03}
In such a case, higher accretion rates are permissible, so that for M87 accretion rates $\dot{m}$ up to some 
$\sim 10^{-4} \dot{m}_{\rm Edd}$ (in terms of the Eddington rate) become possible,\cite{levinson11} values 
that are indeed compatible with the (Bondi) upper limit derived from Chandra observations.\cite{dimatteo03}
Assuming equipartition with the accreted matter, characteristic ADAF magnetic fields in the vicinity ($\sim r_s$) 
of the black hole (and thus likely the maximum magnetic field feeding the jet) would then be of order $B \lppr 
200$ G.\cite{rieger11}\\
(5) M87 has been a prime target for MHD jet simulation studies (see e.g., Refs. \refcite{fendt01}-\refcite{dexter12}).
%\refcite{fendt01,gracia09,broderick09,nakamura10,porth11,dexter12} 
In principle, the maximum possible kinetic power $L_j$ carried by a jet is expected to be comparable to the 
source's accretion power $\sim \dot{m} c^2$ (e.g., Ref.~\refcite{punsly11}). For M87 this translates into an 
upper limit $L_j \lppr 3 \times 10^{44} (\dot{m}/10^{-4}  \dot{m}_{\rm Edd})$ erg/s, which seems consistent 
with other estimates for the time-averaged bulk kinetic power as inferred from its large-scale radio and X-ray 
morphology, ranging from $L_j \sim 3 \times 10^{42}$ up to a few $\times 10^{44}$ erg/s.\cite{owen00,young02,reynolds96b,bicknell96} 
Obviously, if accretion would occur in the standard (cooling-dominated) mode, current accretion rates would 
be too small to account for the mean kinetic jet power.\\ 
(6) Optical (HST) observations of M87 reveal apparent superluminal motion of jet components about $0.5$ 
kpc away from the central black hole, suggesting bulk flow Lorentz factors $\Gamma_j \sim 6$ and a jet 
orientation within $\theta \sim 19^{\circ}$ to the line of sight.\cite{biretta99} Radio (20cm VLBA) observations 
also indicate apparent superluminal speeds up to $\beta_s \simeq 4.3\pm 0.7$ of features moving through 
knot HST-1 (located 0.85 arcsec $\sim60$ pc in projection away).\cite{cheung07} Slower speeds between 
$(0.6-0.8)c$ within 80 pc of the nucleus have been reported based on HST data.\cite{biretta99} 
On the other hand on scales of one parsec ($\lppr1.6$ pc) no significant relativistic motion is evident from 
radio (2cm) observations,\cite{kovalev07} and on sub-parsec scale only mildly relativistic (radio jet) bulk 
speeds ($\beta \lppr 0.7$) and somewhat larger jet inclination angles (yet with velocity measurements based 
on one pair of observations only) have been inferred based on 43 GHz radio VLBI  observations.\cite{ly07}  
A likely range for the jet inclination thus seems to be $\theta \sim (15-25)^{\circ}$.\cite{acciari09Sci} 
This would be largely consistent with M87 being a "non-blazar" source (or "misaligned BL Lacertae 
object"\cite{tsvetanov98}) in the sense that most of its jet emission (at least at radio frequencies) is 
characterized by rather weak Doppler factors $D=1/[\Gamma_j (1-\beta\cos\theta)] \lppr 3$ and moderate 
Doppler beaming only.\cite{dodson06,wang09z}\\
Quite surprisingly, given this seemingly unfavorable condition, M87 has turned out to be a highly variable 
TeV gamma-ray emitter, extending the class of VHE emitting AGN from the classical blazars to radio 
galaxies. Core features at gamma-ray energies are summarized in the following.

\begin{figure}[ph]
\centerline{\epsfig{file=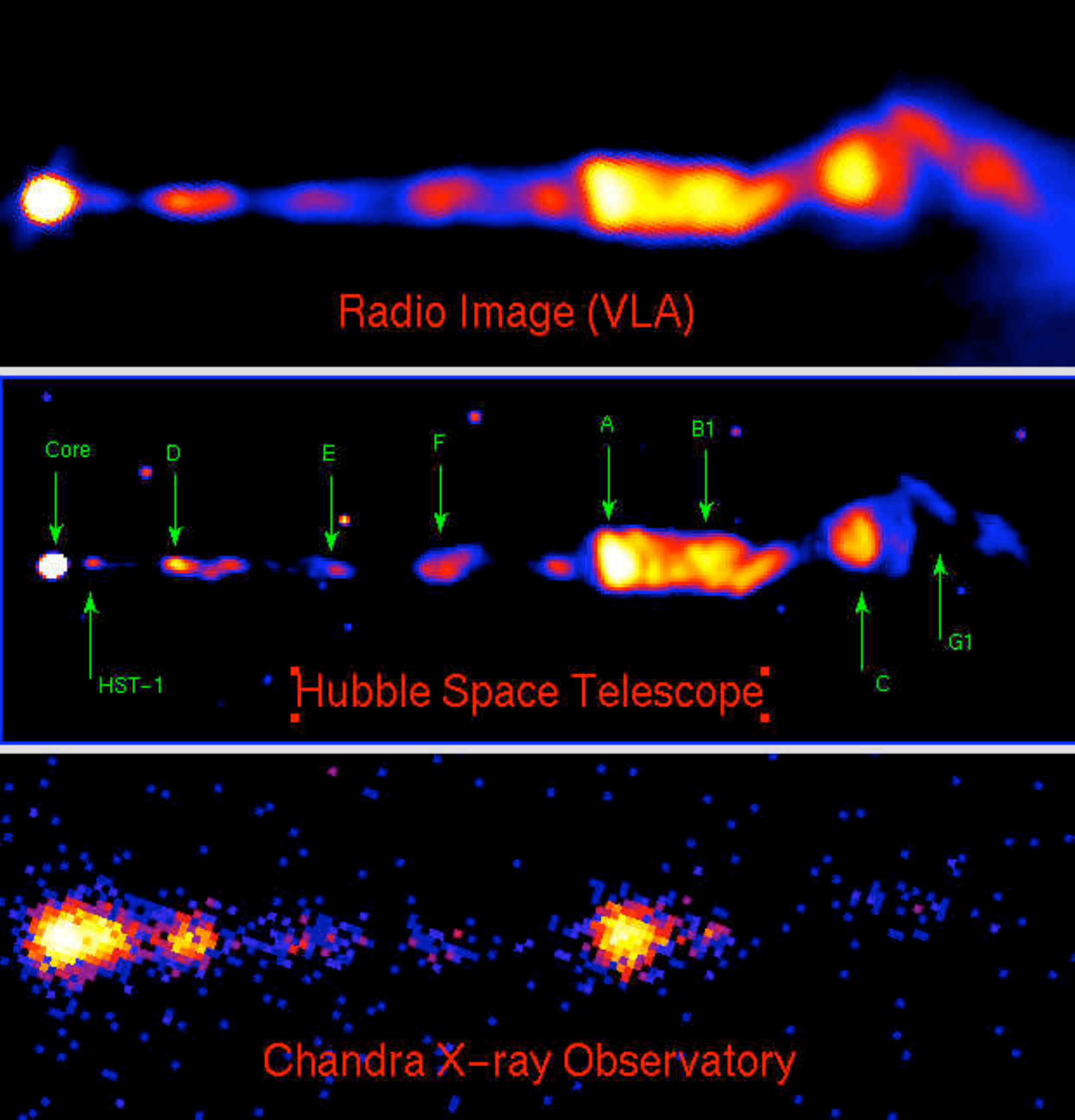,width=3.0in}}
\vspace*{8pt}
\caption{The large-scale jet of M87 as seen in radio, optical and X-rays. The bright knot A is located at about 
$1$ kpc from the nucleus (Credit: X-ray: NASA/CXC/MIT/H.Marshall et al., Radio: F.Zhou, F.Owen (NRAO), 
J.Biretta (STScI), Optical: NASA/STScI/UMBC/E.Perlman et al.)\protect\label{fig1}}
\end{figure}
\begin{figure}[ph]
\centerline{\epsfig{file=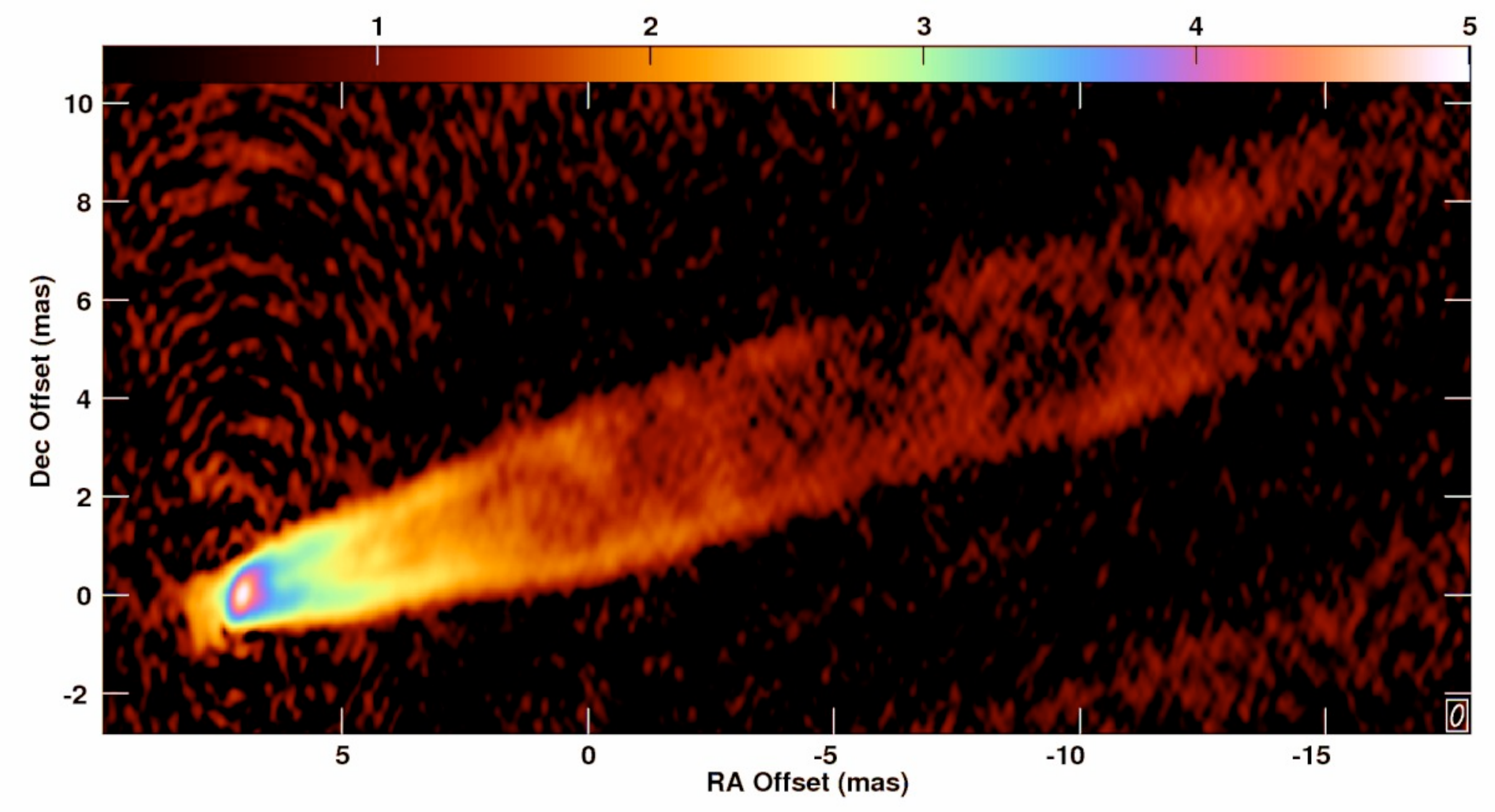,width=5.0in}}
\vspace*{8pt}
\caption{The small (sub-parsec) scale jet in M87 as observed at 43 GHz with VLBA in 2007. The resolution 
achieved is 0.43 by 0.21 mas. Note that for the distance (16.7 Mpc) of M87, 1 mas = 0.081 pc, and 100 
Schwarzschild radii=0.36 mas (for $M_{\rm BH}=3 \times 10^9 M_{\odot}$). (Credit: NRAO: R.C. Walker et 
al., http://www.aoc.nrao.edu/$\sim$cwalker/M87)\protect\label{fig2}}
\end{figure}

\section{Observational findings towards high energies}
\subsection{Very high energy (VHE) results}
\begin{figure}[ph]
\centerline{\epsfig{file=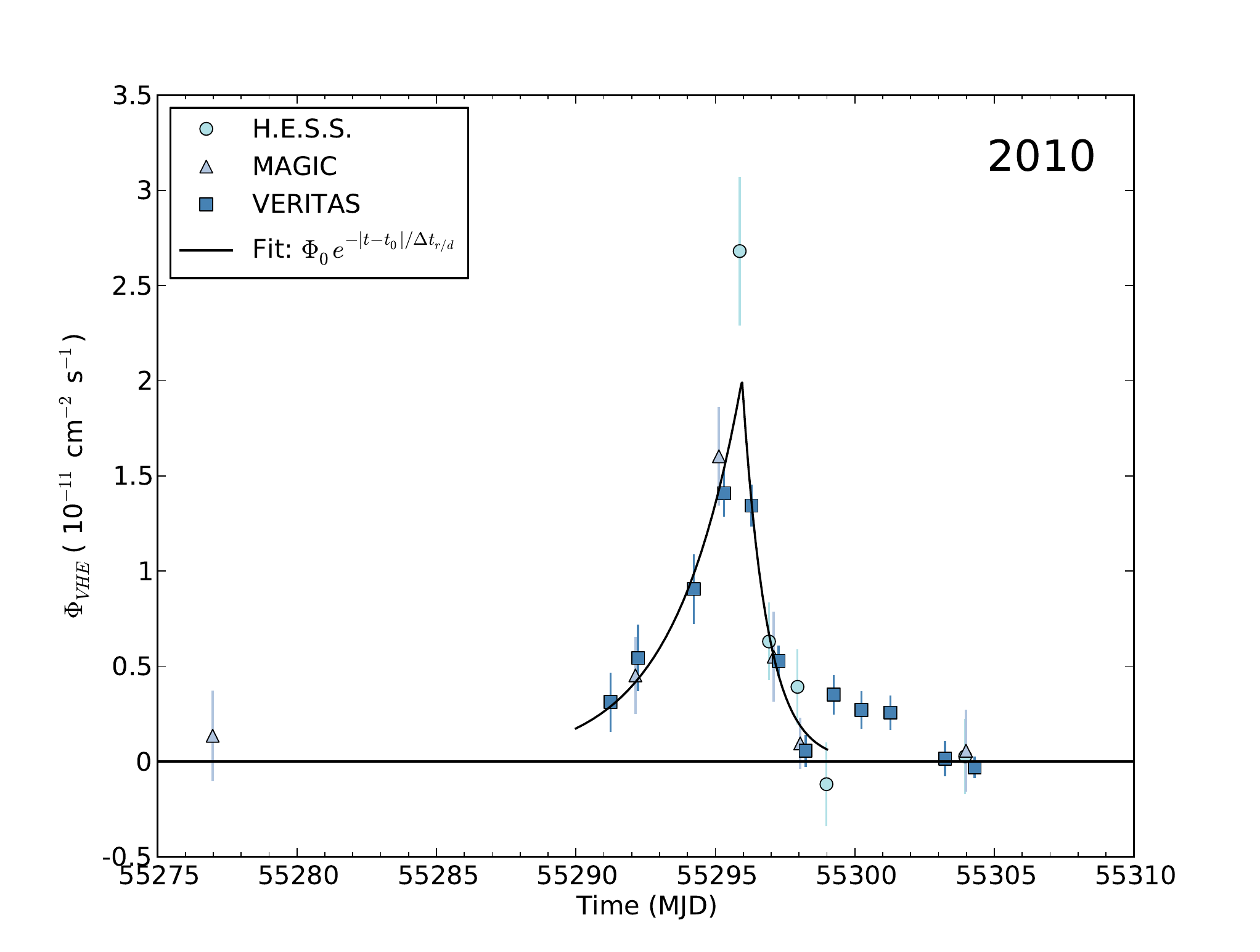,width=4.5in}}
\vspace*{8pt}
\caption{VHE flare of M87 as observed in April 2010 by different instruments. Day-scale activity is 
evident. The curve shows a fit of an exponential function to the data. From Ref.~\protect\refcite{abramowski12}.)\protect\label{fig3}}
\end{figure}

(1) For the first time, a significant ($>4\sigma$) excess of photons from M87 above energies of $\sim700$ 
GeV has been measured by the HEGRA stereoscopic system.\cite{aharonian03} 
Subsequent observations by the current generation of VHE instruments have confirmed gamma-ray 
emission at TeV energies and further demonstrated the exceptional behavior of M87.\cite{aharonian06,acciari08,albert08} \\
(2) H.E.S.S. has detected fast (day-scale) TeV flux variability during a high VHE source state in March/April 
2005 (average photon flux a factor $\sim5$ higher than in 2004), and demonstrated that the VHE flux of 
M87 (between 2003-2006) also varies on long (year-type) time scales.\cite{aharonian06} The measured 
VHE spectrum above 0.35 TeV was found to be consistent with a (single) hard power-law (photon index 
$\Gamma=2.22\pm0.15$, compared to $2.62\pm0.35$ for the 2004 low state), and to extend beyond 
$10$ TeV.
Similar to other wavelengths, the TeV output was relatively moderate, with an average (for the years 2003 
to 2006) isotropic TeV luminosity $L_{\rm TeV}\simeq 3 \times 10^{40}$ erg/s (the luminosity being 
roughly three times larger for the 2005 high state). \\
(3) VERITAS observations during a rather low state in spring 2007 confirmed the hard VHE spectrum and 
provided further evidence for year-scale variability,\cite{acciari08} while MAGIC observations\cite{albert08} 
during another VHE high state in February 2008 again revealed substantial day-scale variability (cf. also 
the 4d-VHE flare observed by VERITAS in 02/2008, Ref.~\refcite{acciari10}). Interestingly, the MAGIC 
findings seemed to indicate a different variability behavior; above 350 GeV considerable flux variability on 
time scales as short as $\sim1$ day has been found, while at lower energies ($150-350$ GeV) no significant 
variability was detected.\cite{albert08} This could perhaps point to the appearance of a new (additional) emission 
component at the highest energies.\\ 
(4) Day-scale VHE activity with a well-defined rise/fall structure of the flux was also observed during another 
VHE high state (exceeding 10\% of the Crab above 350 GeV) in April 2010.\cite{abramowski12,aliu12} VHE 
flux (exponential) rise and decay times of $\sim3$ d and $\sim1$ d, respectively, have been reported by 
VERITAS. See also Fig.~\ref{fig3}.\\
(5) There are indications for spectral variability in this and the previous flares, with photon indices 
(assuming a single power law) in the low (quiescent) and high states closer to $\Gamma\sim 2.6$ and 
$\Gamma\simeq 2.2$, respectively.\cite{aharonian06,albert08,aliu12}\\
(6) Both the hard VHE spectrum extending beyond 10 TeV, and the observed rapid VHE variability down to 
(flux doubling) time scales of $\Delta t_{\rm obs} \simeq 1$ day are remarkable features for a source whose 
overall jet is commonly believed to be substantially inclined to the line of sight. The reported VHE flux 
variations are the fastest observed in any waveband from M87 so far, and already imply a size of the 
$\gamma$-ray emitting region ($R \lppr D \Delta t_{\rm obs} c$), that is of the same order of magnitude as
the Schwarzschild radius of the black hole in M87. Given current evidence, this may possibly point to the 
black hole vicinity as the most likely origin of the VHE radiation.\\
(7) The restricted angular resolution of current Cherenkov telescopes ($\sim 0.1^{\circ}$, corresponding to 
$30$ kpc at the distance of M87) however, does not allow to resolve the jet in M87, so that the true location 
of the VHE region in charge usually remains ambiguous when based on the VHE results alone. On the other
hand, because of weak Doppler boosting, there is not much uncertainty concerning the upper limit on the 
size of the VHE source. In addition, for the 2008 (February) VHE high state a combined radio (VLBA, 43 GHz, 
high-resolution $0.21\times 0.34$ mas) and VHE monitoring effort found that the VHE gamma-ray outburst 
was followed by a strong increase of the radio flux very close to the core (within $\sim 200 r_s$ along the jet 
for a viewing angle $15^{\circ}$; note that there is reasonably experimental evidence that the radio core is 
located close to the black hole, Ref.~\refcite{hada11}). As the flare progressed, the region around the core 
brightened up substantially (associated average $\beta_s=1.1$), eventually reaching radio flux levels beyond 
any seen before, suggesting that a new component was ejected from the core. If the increase of the radio flux 
is caused by a decrease in synchrotron self-absorption, then efficient particle acceleration must take place very 
close to the black hole.\cite{acciari09Sci} Note that a similar radio/TeV connection has not been found for the 
2010 VHE flare.

\subsection{High energy (HE) results}
At high energies ($>100$ MeV), Fermi-LAT reported the detection of gamma-rays from M87 up 
to 30 GeV based on 10 months of data taken between 08/2008-05/2009.\cite{abdo09m87} Given its 
angular resolution ($0.8^{\circ}$ at 1 GeV) the HE location is difficult to pin down. No evidence 
for significant HE flux variations over the period of observations has been found, the reported HE light 
curve (10d-bins) being consistent with no variability. It should be noted, however, that given the low 
HE flux of M87 and the limited sensitivity of Fermi-LAT, a potential variability on smaller (day-scale) 
timescales usually cannot be probed.\\ 
The corresponding isotropic luminosity in the HE band is $L(>100$ MeV) $\simeq 5 \times 10^{41}$ 
erg/s. Interestingly, the HE spectrum essentially appears consistent with a power-law whose photon 
index ($\Gamma=2.26\pm0.13$) is comparable to the VHE one in the high states. However, given the 
observed HE flux, a simple extrapolation of the HE power-law to the VHE regime is not sufficient to 
account for the flux measured during the TeV high states (which, however, only last for a couple of
days and seem to occur with duty cycles $<10\%$). Similar results have also been obtained 
more recently based on an analysis of a larger (two year, 08/2008-08/2010) data set.\cite{abramowski12} 
Comparing the HE fluxes up to 05/2009 with those measured afterwards, no evidence for variability 
in the fluxes above 100 MeV has been found, but there seems to be some weak indications for a 
change in the fluxes above 1 GeV.\cite{abramowski12}

\subsection{X-ray results}
(1) M87 has been a target for several X-ray telescopes, including the Einstein Observatory,\cite{biretta91,schreier82} 
ROSAT\cite{harris97} and XMM-Newton\cite{boehringer01}. However, although the spatial resolutions of these 
instruments were sufficient to detect the jet, observations with the Chandra satellite were needed to resolve the 
jet into more than one knot, and to allow a better localization of the X-ray emission. These observations were 
started in July 2000,\cite{marshall02,wilson02} with a detailed M87 monitoring program implemented in 
2002.\cite{harris03} Given Chandra's subarcsec resolution ($\simeq 0.8$ arcsec), X-ray emission from the 
core (nucleus) and the innermost knot HST-1 can be distinguished.\\ 
(2) The coincidence of VHE $\gamma$-ray flaring activity in 2005 with a giant X-ray flare from HST-1 (with isotropic
X-ray luminosities up to a several $10^{41}$~erg/s) initially prompted speculations that the observed VHE emission 
may have originated in HST-1 as well.\cite{cheung07} The fact, that the Chandra X-ray flux from HST-1 did not 
change much during subsequent VHE flaring episodes (lasting from a few days up to a weak or two) has been 
often taken to disfavor HST-1 as a site of the rapid TeV flaring activity.\cite{abramowski12,acciari08,hardee10} 
It should be noted, however, that from a physical point of view, even a possible anti-correlation is, in itself (!), 
not so much telling, as in a simple inverse Compton model a variation of the magnetic field could lead to 
anti-correlation.\\ 
(3) While Chandra observations during VHE flares have not yet been achieved, the nucleus usually shows 
comparably high X-ray fluxes in Chandra pointing close to the VHE flaring states.\cite{acciari10,harris11}\\
(4) Observed characteristic X-ray variability (flux doubling) time scales for HST-1 are of the order $t_v \sim 0.14$ 
yr ($\sim 50$ d),\cite{perlman03,harris06} while the core (nucleus) seems to display faster variability ($t_{v,c} 
\lppr 20$ d); there is some evidence that the shortest nuclear (doubling) time scale that one can measure with 
Chandra may even be as small as a few days.\cite{abramowski12,harris11,harris09} If true, this would most
likely argue against an ADAF (bremsstrahlung) origin of the Chandra X-ray variations.\cite{hilburn12}
\begin{figure}[ph]
\centerline{\epsfig{file=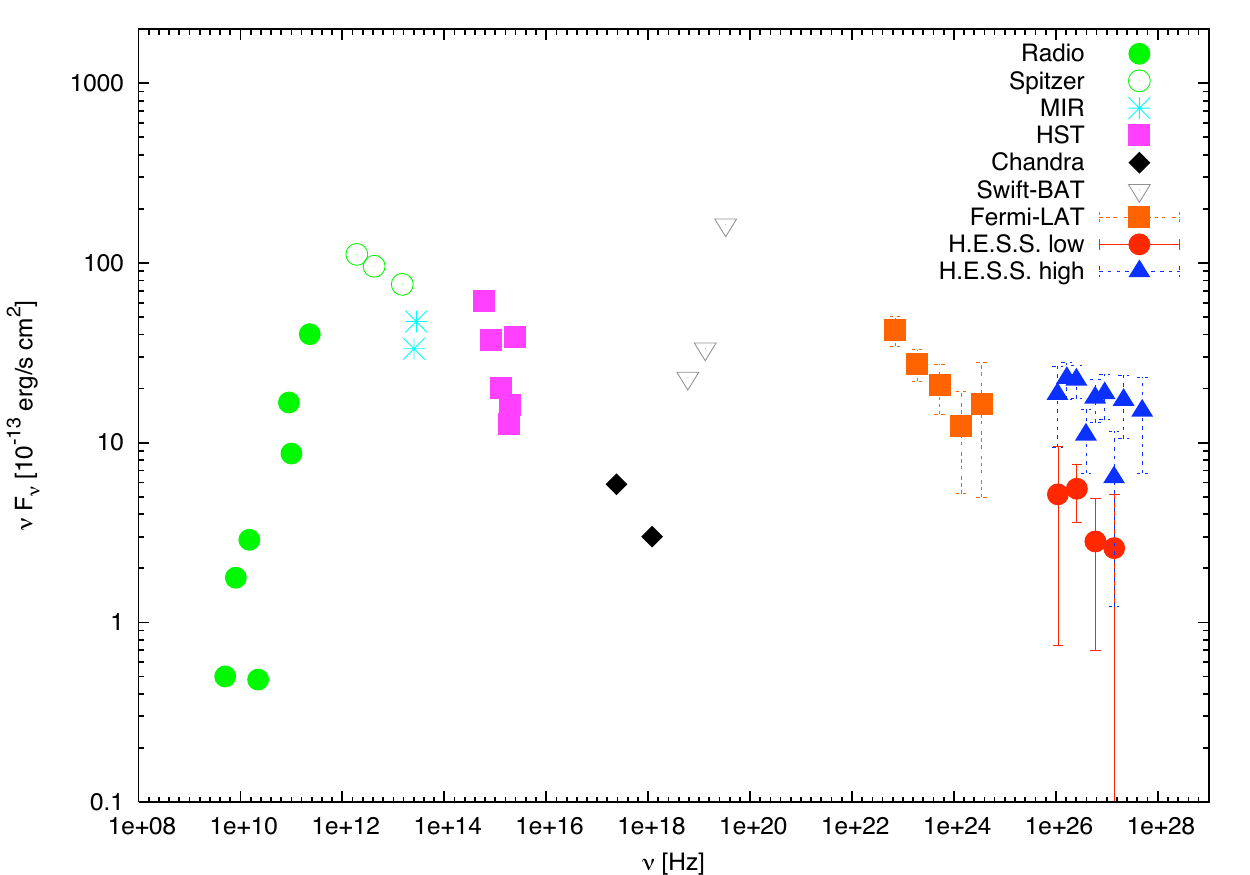,width=4.0in}}
\vspace*{8pt}
\caption{Characteristic (non-simultaneous) nuclear SED of M87. A simple HE (Fermi) power-law extrapolation 
could fit the VHE (H.E.S.S.) low state data, but would under-predict the VHE high state data. Historical data are, 
in increasing order: Radio and sub-mm data [green, filled circles]: 5 GHz VLBI (0.7 mas resolution, Pauliny-Toth 
et al. 1981), VLBA (sub-mas resolution, Zavala \& Taylor 2004), VLBI 22GHz (0.2 mas resolution, Spencer \& 
Junor 1986), IRAM/89 GHz (2 arcsec resolution only, upper limit for nuclear flux, Despringre et al. 1996), 
VLBI/100 GHz (0.1 mas resolution, Baath et al. 1992), SMA 230 GHz (300 mas resolution, Tan et al. 2008). 
Spitzer FIR data [green, open circles] (several arcsec resolution, Shi et al. 2007). MIR data (cyan, star) from 
Keck ($\sim300$ mas resolution, Whysong \& Antonucci 2004) and Gemini (460 mas resolution, Perlman et 
al. 2001). Optical HST data (pink, squares) are from Sparks et al.~1996 ($\sim200$ mas resolution). Nuclear 
X-ray data (black, diamonds) are from Chandra observations ($\sim500$ mas resolution, Marshall et al. 2002) 
in July 2000. Note, however, that the nuclear X-ray flux may increase by a factor $\sim (5-10)$ during strong flaring 
states (as e.g. in February 2008, Harris et al. 2009). Swift/BAT hard X-ray longterm (03/2005-01/2009) upper 
limits [grey, open triangles] (Ajello et al. 2009; Abdo et al. 2009). Fermi-LAT longterm HE data (08/2008-05/2009) 
[orange, filled squares] (Abdo et al. 2009). VHE data from H.E.S.S. for the 2005 flaring (blue, triangles) and 2004 
low state (red, filled circles) (Aharonian et al. 2006).)\protect\label{fig4}}
\end{figure}

\section{Theoretical interpretations}
A variety of models have been proposed in the literature to account for the HE and VHE $\gamma$-ray emission 
from M87, with production sites ranging from large (hundred pc and beyond; e.g., knot HST-1 or knot A) via the 
innermost (sub-pc) jet scales down to the immediate environment (magnetosphere) of the supermassive black 
hole. Some representative candidate sites and mechanisms will be discussed in the following.

\subsection{HST-1}
The production of TeV photons by inverse Compton (IC) scattering requires TeV electrons. As these electrons also 
emit optical to X-ray synchrotron emission, the bright optical/X-ray features in M87 are putative sites for the origin 
of the observed TeV emission. A priori, this includes the inner jet (core), the HST-1 complex (at a de-projected distance 
$z>120$ pc) and knot A (at a distance of about  $1$ kpc). However, knot A, with a typical resolved X-ray size $\sim 80$ 
pc,\cite{wilson02} is commonly considered to be disfavored (based on light travel arguments)\cite{aharonian06} as 
a significant contributor to the observed variable VHE emission on daily and yearly time scales. HST-1 (typical radio 
size $\sim0.2$ pc),\cite{harris09} on the other hand, has been thought to remain a promising site where IC up-scattering 
of ambient starlight photons by shock-accelerated electrons could lead to variable TeV emission.\cite{cheung07,stawarz06} 
This complex has shown interesting activities during the past years, including phases of strong optical and X-ray 
brightening and the appearance of superluminal sub-structures.\cite{cheung07,giroletti12,harris06}\\ 
The fact that there seems to be no longterm correlation between the Chandra HST-1 X-ray and the TeV flux does not 
necessarily (as argued above) provide cogent evidence against such a scenario. Instead, the observed rapid TeV 
variability on time scale of $\sim1$ day, implying a source size of $R \lppr D c \Delta t \sim 8 \times 10^{15}(\Delta 
t/1~\rm{d})(D/3)$ cm, appears much more challenging to account for. In the case of a conical jet structure, for example, 
an unlikely small opening angle $\theta_j = R/z \lppr 2\times 10^{-5}$rad would be required.\cite{bromberg09} 
Perhaps more likely is a scenario where HST-1 is the site of a hydrodynamic re-collimation shock (e.g., Ref.~\refcite{cheung07},
but see also Ref. \refcite{asada12})\cite{tsinganos02} where convergence occurs and part of the jet kinetic energy is 
converted into radiative energy (e.g., X-rays and possibly $\gamma$-rays). In order to account for the observed fast 
(days-scale) TeV variability, strong collimation would be needed. Recent calculations suggest, however, that due to 
its low radiative efficiency, only modest collimation is achievable. While sufficient to account for the X-ray variability 
observed from HST-1 ($t_v \sim 0.14$ yr), it would not be sufficient to account for day-scale TeV variability.\cite{bromberg09} 
Moreover, the radio to X-ray spectrum of HST-1 (in 2005) can be successfully modeled as a broad-band synchrotron 
spectrum with a cooling break at $\nu_b \sim 10^{15}$ Hz.\cite{harris06,rieger08b} Plausible value for the magnetic 
field strength at the position of HST-1 are of the order $5 \times 10^{-3}$G and below, suggesting a characteristic 
synchrotron cooling time scale for electrons up-scattering starlight photons to the TeV regime of $t_s' \sim (10^6/
\gamma')(5\times10^{-3}/B')^2$ yr. This value is in fact close to the observed variability time scale in the UV.\cite{harris09}
It seems hard to imagine that any supposed short-term TeV variability of HST-1 will not be smeared out on 
comparable time scales. Based on current evidence, it thus seems that (at least) for the episodes of rapid TeV 
flaring activity an origin in the sub-parsec scale jet and below is preferred. It could well be, however, that HST-1, the
kpc-scale jet\cite{hardcastle11} and/or perhaps even knot A\cite{honda07} contribute to the overall, quiescent VHE 
emission in M87.

\subsection{Leptonic (inner) jet models}
In this approach, the (non-simultaneous) SED representation is often interpreted as showing a break between 
$10^{12}-10^{13}$ Hz, and a possible synchrotron cut-off around or below $10^{18}$ Hz. If this is indeed true, 
however, then homogeneous (one-zone) synchrotron self-Compton models are (under the proviso of moderate 
Doppler boosting) at best only marginally able to match the quiescent TeV flux in M87 and generally unable to 
account for its flaring state.\cite{abdo09m87,georg05,tavecchio08}\\ 
A number of different alternatives have thus been considered that could allow accommodating the TeV results. 
These include: Dropping the single zone approach by assuming (1) a longitudinal jet stratification (as in the 
decelerating jet flow/upstream Compton [UC] scattering scenario, Refs. \refcite{georg03,georg05}), (2) a radial 
jet stratification (as in the spine-shear approach)\cite{tavecchio08} or (3) multiple emission zones with changing 
line-of-sight (LOS) directions (differential Doppler boosting).\cite{lenain08,giannios10}\\ 
In models of the type (1),\cite{georg05} the TeV emission turns out to be dominated by inverse Compton emission 
from the fastest ($\Gamma_j$ up to a few tens) part of the jet on target (synchrotron) photons mainly produced in 
the downstream (further away and slower, $\Gamma_j \sim$ a few) jet parts. While such models bear the potential 
to resolve some BL Lac/FR~I unification problems, they are, when applied to M87, not without their challenges: 
The resultant VHE spectra tend to be too soft; there is as yet no direct observational evidence for a decelerating 
flow, in fact, the jet in M87 rather seems to be accelerating (see 2.~(6) above); moreover, the small Doppler 
factor (e.g., $D =1.5$ for $\Gamma_j=10$ and $\theta =20^{\circ}$) associated with the fastest part of the jet and 
the anticipated scale of the jet base usually makes it quite challenging to accommodate the observed short time 
(day-scale) TeV variability.\\ 
Models of the type (2), on the other hand, rely on a possible radiative interplay between a fast inner core (spine) 
and a slower outer (shear) layer. The underlying assumption of a velocity gradient across the (inner) jet is indeed
compatible with radio VLBA images\cite{kovalev07} showing a limb-brightened structure (cf. also Refs. 
\refcite{dodson06,matvey11}) and seems supported by relativistic jet simulations.\cite{hardee10} 
In principle, such a setup is known to be favorable to further shear-type particle acceleration.\cite{rieger04} 
Yet, even when this is neglected, each component should see the (external) radiation of the other boosted, thereby 
enhancing the inverse-Compton emission of both the spine and the layer. Given the supposed inclination of the 
jet in M87, its infrared-optical spectrum is expected to be dominated by synchrotron emission from the weakly-beamed 
(!) spine and its TeV part by (external) inverse Compton emission from the stronger-beamed layer.\cite{tavecchio08} 
The scenario requires the zones to be (quasi) co-spatial for the radiative interplay to work efficiently. This, however, 
could also lead to some problems concerning the observed TeV emission in that a simultaneous fit to the broadband 
spectrum seems to imply a rather dense (weakly de-beamed) infrared-optical radiation field resulting in considerable 
$\gamma\gamma$-absorption and a steep slope at TeV energies. Existing model calculations in fact appear unable,
for example, to reproduce the hard (2005) TeV spectrum seen by H.E.S.S., and it seems not yet obvious how this 
might best be recovered.\\ 
In ("mini-blob") models of the type (3), small sub-regions (e.g., plasma blobs) are taken to follow different trajectories 
within the jet formation zone ($\sim 50 r_s$), thereby allowing for situations where the velocity vector of a region 
becomes close to the line of sight, so that for this region strong Doppler boosting effects become possible (despite 
the overall misalignment). Such a scenario could in principle account for VHE high states that are accompanied by 
a dramatic change in X-ray flux (as related to a SSC-emitting blob moving along the LOS). Again, the success is not 
without its challenge. Modeling of the VHE high state in 2005,\cite{lenain08} for example, indicates a significant deviation 
from equipartition, with the pressure in relativistic electrons exceeding the one in magnetic field by a factor $\gg 10^3$. 
Moreover, required magnetic field strengths in the emitting region ($B\sim 10$ mG) would be a factor $\sim 100$ lower 
than the value anticipated for the local (jet) field based on magnetic flux conservation. Perhaps, some of the restrictions 
could be relaxed if these blobs are also considered to carry a significant azimuthal velocity component as expected 
for a rotating jet flow near its base,\cite{camenzind92,rieger04} or if the blobs (also called "mini-jets" to distinguish 
them from the above) are allowed to have an additional relativistic velocity ($\Gamma_{c}\gg1$) relative to the mean 
jet flow (characterized by $\Gamma_j$). 
The latter situation may be the outcome of relativistic (Petschek-type) magnetic reconnection events\cite{lyubarsky05} 
occurring in a highly magnetically-dominated (initial $\sigma \gg 1$, i.e., magnetic energy $\gg$ rest mass energy of 
the matter) electron-proton jet.\cite{giannios10,giannios09} If the fields can be efficiently dissipated (downstream 
$\sigma_e <1$), with strong mini-jets developing at distances of $\sim 100 r_g$ and reaching $\Gamma_c \simeq 
\sqrt{\sigma} =10$, up-scattering of external nuclear photons (e.g., from an ADAF, see Ref.~\refcite{cui12}) could 
account for the observed TeV emission. Such a model is interesting in that it allows for strong Doppler beaming effects 
and seems capable of accommodating even shorter VHE variability timescale.\cite{giannios10} On the other hand, 
it is not clear whether the assumed conditions are likely to prevail: Characteristic magnetization parameters for 
electron-proton (disk-driven) jets, for example, are usually believed to be much smaller ($\sigma \ll 100)$\cite{camenzind99}; 
whether/how efficient (non-thermal) power-law electron acceleration much beyond the thermal Lorentz factor $\sqrt{\sigma}
m_p/2m_e \sim 10^4$ can be achieved, still remains to be understood. Moreover, for the assumed conical jet geometry 
(opening angle $\sim 1/\Gamma_j$, but see 2.(3) above), the dynamical role of the anticipated guide-field in the jet 
frame at $r\sim 100 r_s$ is likely to be non-negligible (i.e., the reconnecting fields are not strictly anti-parallel), in which 
case the magnetization of the outflow ($\sigma_e>1$) could be still high, thereby leading to only weak and slow mini-jets.

\subsection{Hadronic jet models}
Hadronic interactions could in principle also contribute to or even dominate the observed VHE emission in M87.\\ 
(1) A hadronic inner jet model including synchrotron emission of energetic primary protons as well as the 
contribution of secondary charged pions (from photo-meson $p\gamma$-production) and muons (from pion 
decay) has been put forward a while ago to describe its time-averaged spectral energy distribution.\cite{reimer04} 
If one assumes proton synchrotron emission to dominate, maximum (co-moving) proton energies (assuming 
shock-type acceleration with $t_{\rm acc}'=\eta r_g'/c$ and efficiency $\eta\geq 1$) would be limited to $\gamma_p'
\simeq 4\times 10^{10} \eta^{-1/2} (30~\rm{G}/B')^{1/2}$. The resultant observed synchrotron emission would 
then essentially be constrained to energies $h\nu_p \lppr 0.3 D/\eta$ TeV (above which the spectrum decays 
exponentially). Hence, if M87 would be of the misaligned blazar-type (with small Doppler factor $D$), one would 
not expect its hard TeV spectrum to be accounted for by proton synchrotron emission. The mean pion energy 
from photo-meson production, on the other hand, is (in the delta functional formalism) $\gamma_{\pi}' \simeq 0.2 
\gamma_p' m_p/m_{\pi}$, so that $h\nu_{\mu} \simeq 28 h\nu_p \lppr 8 D/\eta$ TeV. This process could thus 
produce VHE emission up to $\sim 10$ TeV if protons are able to reach maximum Lorentz factors $\gamma_p'$, 
i.e., provided extreme acceleration $\eta \sim 1$ would indeed occur. As $\eta \propto (c/u_s)^2$, this would 
require relativistic shock speeds beyond what has been seen so far. It seems not clear yet, whether such a 
model could self-consistently produce the hard TeV spectrum. Photon-meson production in M87 is expected 
to take place on characteristic (co-moving) time scales $t_c'\gppr 1.7 d~(L_{\gamma}'/5\times10^{41}
\rm{erg/s})(5\times10^{15}\rm{cm}/R')^2 (\epsilon_{\gamma}/0.01 \rm{eV})$, so that the observed variability
could perhaps be compatible with current constraints.\\  
(2) Alternatively, VHE $\gamma$-rays may originate in inelastic proton-proton collisions (with characteristic 
cooling time scale $t_{pp}' \simeq 10^{15}/n_p'$ sec). The required high target proton density ($n_p'$) needed 
to account for day-scale VHE activity could possibly be introduced by a star or a massive, dense gas cloud (of 
typical initial radius $r_{c0}\sim 10^{13}$ cm and mass $M_c \sim 10^{29}$ g) penetrating into the jet on scales 
of some tens of Schwarzschild radii.\cite{barkov10,bosch12,barkov12} If the jet is powerful enough and if a sufficiently 
large fraction ($\eta \geq 0.1$) of the power tapped by a cloud can be channeled into efficient acceleration of 
protons, such a scenario could well explain hard and variable VHE emission from non-blazar AGN. For an 
almost collimated jet (opening angle $\theta \sim 2^{\circ}$) detailed spectral modeling of the M87 VHE flare 
in April 2010 implies a jet power at the upper end of the estimates presented above (cf. 2.(5)).\cite{barkov12} 
Assuming that to be the case, the model can successfully explain the observed VHE (flare) light curve and 
spectrum. To accommodate the observed, large jet opening angle (and thus the jet radius $r_j$, cf. 2.(3)), 
this approach requires a situation where most of the jet energy flux is still concentrated into a narrow core 
(spine) as otherwise the characteristic expansion time scale of the cloud (determining the variability) might 
become too large and the jet power tapped by the cloud ($\propto (r_c/r_j)^2$) too small to account for the 
observed VHE characteristics. It seems conceivable that such a situation could arise in a hybrid black 
hole/disk-driven jet scenario.\\  
(3) A combined lepto-hadronic jet model (including proton synchrotron and proton-proton interactions as 
well as energetic particle transport by convection) has been analyzed recently,\cite{reynoso11} showing 
that the time-averaged nuclear broadband spectrum of M87 might be reasonably accounted for (with its 
high energy SED part dominated by pp-processes), if the kinetic jet power would indeed be able to  
approach (perhaps occasionally) values as high as $L_j \sim 10^{46}$ erg/s. This seems certainly a strong
condition when compared with current time-averaged estimates for the jet power (cf. 2.(5)).\\
A unique signature of a hadronic jet model would be the detection of high energy neutrinos. However, 
typical model predictions lie a few orders of magnitude below current instrumental sensitivity.

\subsection{Magnetospheric VHE emission}
The observation that the VHE emission varies on characteristic timescales comparable to the light travel
time across (at most) a few Schwarzschild radii, and the finding that the VHE flaring activity in 2008 was   
accompanied by an increase in radio flux close to core, suggest that the relevant non-thermal particle
acceleration and emission processes could in fact take place in the very vicinity of the central black hole 
itself.\cite{acciari09Sci} Magnetospheric models, where these processes are considered to occur at the 
base of a rotating black-hole-jet magnetosphere,\cite{levinson11,rieger08a,levinson00,neronov07,rieger08b} 
may then become particularly interesting. In principle, at least two main variants can be distinguished 
(see Ref.~\refcite{rieger11} for an extended review):\\
(1) Particle acceleration in magnetospheric vacuum gaps (direct electric field acceleration), with primary
electrons in M87 injected by, e.g., annihilation of ADAF disk (MeV) photons and quickly reaching high  
Lorentz factors ($\gamma_e \sim 10^{10}$).\cite{levinson11} Curvature emission and inverse Compton 
(IC) up-scattering of ADAF soft photons can then produce VHE photons with a spectrum extending up 
to $10^4$ TeV. VHE photons with energies below some TeV might be able to escape absorption (see 
below), while IC photons with energies well above 10 TeV are expected to interact with the ambient 
radiation field and initiate pair cascades (e.g., Refs.~\refcite{neronov07,vincent10}) just above the gap, 
leading to a large multiplicity.\cite{levinson11} The results suggest that this could ensure a force-free 
outflow above the gap and account for the appearance of the VLBA jet. Intermittencies of the cascade 
process, induced by, e.g., modest changes in accretion would give rise to the variability of the TeV 
emission and to fluctuations of the resultant force-free outflow. The HE and VHE parts of the spectrum 
are then likely to be shaped by different emission processes. Models of this type depend on a sufficiently 
spinning black hole to power the VHE emission, require the presence of charge-starved regions in the 
magnetosphere and rely on a radiatively inefficient disk configuration to ensure escape of TeV photons. 
Significant $\gamma\gamma$-absorption features should become apparent in the VHE spectrum above 
$\sim 10$ TeV.\\
(2) If the plasma density is such that the parallel electric field becomes effectively screened, inertial 
(centrifugal) effects could probably still allow efficient acceleration of electrons to energies of several 
TeV. The general picture would then be one where energetic particles experience the centrifugal force 
while moving along rotating magnetic field lines anchored in the innermost part of the accretion disk, 
with electrons reaching suitable high Lorentz factors for IC up-scattering of ambient ADAF soft photons 
to the VHE regime close to the light cylinder $r_{\rm L}$.\cite{rieger08b} Models of this type still need 
a sufficiently spinning black hole (Kerr parameter $\gppr 0.8$) for $r_{\rm L}$ to be small enough to 
account for day-scale VHE activity. In addition, any encountered random magnetic fields should be 
small enough for electron synchrotron emission not to exceed current HE constraints.\cite{rieger08a,neronov07}

\subsection{Escape of VHE $\gamma$-rays?}
An important issue for inner jet scenarios, and in particular magnetospheric VHE models, concerns the 
question whether TeV photons are able to escape photon-photon interaction ($\gamma\gamma$-absorption) 
in the source's vicinity. In principle, TeV photons of energy $\epsilon$ interact (producing pairs) most 
efficiently with target photons in the infrared regime $\epsilon_T \simeq (1\,\rm{TeV}/\epsilon)$ eV. Escape 
would thus only be possible if the infrared (IR) photon field is weak or sufficiently diluted. This could happen 
if the observed IR field is dominated by e.g. (anisotropic) synchrotron jet emission and/or a dusty torus on 
larger scales. For M87, ground-based (Keck LWS, 300 mas resolution) mid-infrared (MIR) observations at 
$0.1$ eV suggest (under the assumption of isotropy!) a nuclear luminosity $L_{\rm MIR} \simeq 10^{41}$ 
erg/s within $20$ pc from the black hole (cf. also Gemini/460 mas resolution, and VLT-VISIR/350 mas resolution, 
MIR observations for similar results\cite{perlman01,reunanen10,asmus11}).\cite{whysong04} The MIR findings 
are broadly consistent with a non-thermal synchrotron jet origin of the nuclear IR emission.\cite{perlman01} 
Along with other, more recent IR observations (Spitzer MIR $1.5-6 \times 10^{13}$ Hz, $\sim3''$ resolution; 
Herschel FIR $0.6-3 \times 10^{12}$ Hz, $\sim8''$ resolution), they seem to disfavor a significant/dominant 
(quasi-isotropic) IR contribution by thermal dust (torus) emission in M87 (but see also Ref.~\refcite{perlman07} 
for a possible dust contribution towards longer wavelengths).\cite{buson09,baes10,mason12} 
Experimentally, this would imply an optical depth for TeV photons $\tau_{\gamma\gamma}(\epsilon)\simeq 
\sigma_{\gamma\gamma} n_s(\epsilon_T) R_d=0.05 \eta (L_T/10^{41}\rm{erg/s})(0.1\rm{pc}/R_d)(\epsilon/
\rm{1 TeV})$, considering the IR contribution to be dominated by jet emission (with associated anisotropy 
$\eta\ll1$) on scales $R_d\lppr 0.1$ pc. Under such conditions TeV photons may well escape absorption. 
Alternatively, one may calculate the expected IR photon field in the vicinity of the black hole by specifying 
some inner disk configuration and accretion rate, and then use this to derive optical depths. If the inner disk 
would be of the standard (i.e., cooling-dominated, radiatively efficient) type, the emergent disk photon field 
would peak in the IR and be dominated by emission from regions close to the marginal stable orbit, thus 
making an escape of TeV photons from regions close to the black hole rather unlikely.\cite{rieger11,brodatzki11} 
Most probably, however, the (inner) accretion flow in M87 is radiatively inefficient and dominated by advection 
(ADAF, see above 2.(4)). In such a case, escape of TeV $\gamma$-rays would again be feasible.\cite{rieger08b} 
Even if, for example, the whole nuclear radio to X-ray flux observed in M87 would solely originate in an ADAF 
(which is certainly over-restrictive given that most of its nuclear radio-X-ray emission is usually believed to be 
jet-dominated, e.g. Ref.~\refcite{perlman01}), then in the presence of a rapidly rotating back hole $\sim 10$ 
TeV photons would be able to escape absorption once produced on scales $\gppr 5 r_s$.\cite{wang08} 
Detailed calculations show that the anticipated optical depth for TeV photons is sensitively dependent on 
accretion rate,\cite{levinson11,li09} so that magnetospheric VHE emission (provided it indeed occurs) may 
in fact be modulated by small changes in accretion conditions. Given current evidence there are thus good 
reasons to believe that, at least for M87, TeV $\gamma$-rays may be able to escape from the vicinity of its 
supermassive black hole. This in turn suggests that future high-sensitivity instruments such as the European 
CTA project could play a particularly important role in testing magnetospheric VHE $\gamma$-ray emission 
scenarios in M87.

\subsection{On dark matter, ultra-high-energy cosmic rays and EBL prospects}
The observed VHE variability excludes a dominant gamma-ray contribution due to cosmic-ray interactions with 
the thermal gas in the interstellar medium\cite{pfrommer03} or neutralino annihilation inside the dark matter 
halo of M87,\cite{baltz00} although it remains possible that some of these processes could contribute to its low 
state (long-term average) gamma-ray emission.\cite{saxena11}\\
The relevance of M87 as candidate source for the origin of ultra-high-energy cosmic rays (UHECR, $>4 
\times 10^{19}$ eV) has often been discussed in the literature, see e.g. Ref.~\refcite{farrar00,boldt00,protheroe03,honda04}. 
Efficient proton acceleration to ultra-high energies in magnetospheric vacuum gaps, however, appears rather
unlikely (due to curvature losses and partial field screening).\cite{levinson11,rieger11} On the other hand, 
diffusive shock acceleration of UHE protons in the jet of M87 (cf. also the proton synchrotron model described 
above) seems marginally possible provided relativistic shock speeds would indeed occur. Given its proximity 
and anticipated relatively small deflection angles (for protons), M87 is one of a few potential extragalactic 
sources that could possibly allow to test UHE cosmic-ray acceleration by directional $\gppr 10^{19}$ eV 
cosmic-ray astronomy.\cite{sommers09,bauleo09}\\
The H.E.S.S. observations in 2005 indicate that the VHE spectrum of M87 could well reach up to a few tens of 
TeV. Due to its proximity, the VHE $\gamma$-ray flux of M87 is likely not to become significantly affected by 
interactions (pair production) of TeV photons with infrared photons of the diffuse extragalactic background light 
(EBL), at least  up to some tens of TeV. An extension of the hard power-law $\gamma$-ray spectrum of M87 
beyond 30 TeV, for example, would thus allow a very unique probe of the EBL in the transition region from 
mid- to far infrared ($\lambda \sim30~[\epsilon_{\gamma}/\rm{30~TeV}] \mu$m).\cite{aharonian04book} 
The measured TeV fluxes and anticipated sensitivities of the future Cherenkov array CTA could make this 
a prime physics case.

\section{Conclusions}
The radio galaxy M87 is a unique source on the extragalactic sky. Its proximity and huge central black hole 
mass allow for an analysis of fundamental astrophysical processes in unprecedented detail. M87 is well-known 
for a one-sided, kpc-size relativistic jet visible at radio, optical and X-ray wavelengths. However, compared to 
typical blazar sources, this jet is most likely substantially inclined $\sim(15-25)^{\circ}$ with respect to the 
observer, so that the broadband emission observed from M87 is not necessarily (everywhere) swamped by 
relativistically-beamed jet radiation. M87 thus appears to be a special active galaxy where we may be able 
to discover central features related to other high-energy processes. The rapid (days-scale) VHE flux variability 
detected during high source states by current $\gamma$-ray instruments is the fastest variability seen so far 
at any wavelength, and suggests that the variable, hard-spectrum $\gamma$-ray emission may originate 
close to its supermassive black hole. A variety of physical scenarios have been proposed in the literature to 
accommodate such findings. 
Some of the pros and cons/limitations of these candidate models are mentioned in this review, and the relevance 
of further VHE observations is stressed. VHE flaring characteristics (energy spectra and extension, variability 
time scales) like those reported bear the rare potential to give us important insights into the near black hole 
environment of active galaxies. Future, more sensitive VHE observations (CTA) will allow to study spectral 
variability and further constrain the time scales of VHE flux variations - both important inputs for theoretical
modeling. Simultaneous VHE/high-resolution radio observations, on the other hand, will have the potential 
to pin down the location of the VHE emission on spatial scales $\lppr 10^2 r_s$. M87 is the best-suited AGN 
for such kind of studies, since it is (beyond Sgr~A*) the most promising object for which direct imprints of the 
black hole in future high-resolution radio observations can be expected.\cite{dexter12}

\section*{Acknowledgments}
We are grateful to Amir Levinson, Maxim Barkov and Matthias Beilicke for helpful comments on the manuscript.

\end{document}